\documentclass{aa}

\usepackage{graphicx}
\usepackage{natbib}
\usepackage{txfonts}
\usepackage{amssymb}
\usepackage{threeparttable}
\bibpunct{(}{)}{;}{a}{}{,} 
\begin{document}

   \title{Observations of white-light flares in NOAA active region 11515: high occurrence rate and relationship with magnetic transients }
   \author{Y. L. Song\inst{1,2}
           \and
           H. Tian\inst{1}
           \and
           M. Zhang\inst{3,4}
           \and
           M. D. Ding\inst{5}
          }
   \offprints{H. Tian}
   \institute{School of Earth and Space Sciences, Peking University, Beijing 100871, China; \\
             \email{huitian@pku.edu.cn}
             \and
             State Key Laboratory of Space Weather, Chinese Academy of Sciences, Beijing 100190, China
             \and
             Key Laboratory of Solar Activity, National Astronomical Observatories, Chinese Academy of Sciences, Beijing 100012, China
              \and
              School of Astronomy and Space Science, University of Chinese Academy of Sciences, Beijing 100049, China
              \and
              School of Astronomy and Space Science, Nanjing University, Nanjing 210093, China
             }
   \date{}

\abstract
{}
{There are two goals in this study. One is to investigate how frequently white-light flares (WLFs) occur in a flare-productive active region (NOAA active region 11515). The other is to investigate the relationship between WLFs and magnetic transients (MTs).}
{We use the high-cadence (45s) full-disk continuum filtergrams and line-of-sight magnetograms taken by the Helioseismic and Magnetic Imager (HMI) on board the Solar Dynamics Observatory (SDO) to identify WLFs and MTs, respectively. Images taken by the Atmospheric Imaging Assembly (AIA) on board SDO are also used to show the morphology of the flares in the upper atmosphere. }
{We found at least 20 WLFs out of a total of 70 flares above C class (28.6\%) in NOAA active region 11515 during its passage across the solar disk ($E45^\circ \sim W45^\circ$). Each of these WLFs occurred in a small region, with a short duration of about 5 minutes. The enhancement of white-light continuum intensity is usually small, with an average enhancement of 8.1\%. The 20 WLFs observed were found along an unusual configuration of the
magnetic field characterized by a narrow ribbon of negative field. Furthermore, the WLFs were found to be accompanied by MTs, with radical changes in magnetic field strength (or even a sign reversal) observed during the flare. In contrast, there is no obvious signature of MTs in those 50 flares without white-light enhancements.}
{Our results suggest that WLFs occur much more frequently than what was previously thought, with most WLFs being fairly weak enhancements. This may explain why WLFs are not frequently reported. Our observations also suggest that MTs and WLFs are closely related and appear co-spatial and co-temporal,
when considering HMI data. A larger enhancement of WL emission is often accompanied by a larger change of the line-of-sight component of the unsigned magnetic field. Considering the close relationship between MTs and WLFs, many previously reported flares with MTs may be WLFs.}

\keywords{Sun: activity-Sun: chromosphere-Sun: photosphere-Sun: flares-Sun: magnetic fields}

\titlerunning{White-light flares in NOAA AR 11515}
\authorrunning{Song et al.}
\maketitle

\section{Introduction}

White-light flares (WLFs) are defined as flares with a sudden enhancement of emission in the optical continuum ({\v S}vestka 1970; Neidig 1989), which are often believed to be very rare (\v{S}vestka 1966; Neidig \& Cliver 1983; Fang et al. 2013). Since the observation of the first WLF in 1859 (Carrington 1859, Hodgson 1859), the number of WLFs recorded in literature is very small compared to the total number of solar flares. Only about 150 WLFs have been reported conclusively in the literature, up to the beginning of this century (Fang et al. 2013). Though the number of observed WLFs is small, they are important because they challenge our knowledge on the transportation of flare energy (Neidig 1989) and the heating mechanisms of the lower solar atmosphere (Ding et al. 1999a). To understand WLFs, several heating mechanisms have been proposed: electron beam bombardment  (Hudson 1972, Aboudarham \& H\'enoux 1986), soft-X-ray irradiation (H\'enoux \& Nakagawa 1977), Alfv\'en wave dissipation (Emslie \& Sturrock 1982, Fletcher \& Hudson 2008), backwarming (Machado et al. 1989, Metcalf et al. 1990, Heinzel \& Kleint 2014) and chromospheric condensation (Gan et al. 1994, Kowalski et al. 2015b).

Based on different characteristics in observations, WLFs are classified into two types (Machado et al. 1986). For type I WLFs,  there is a strong correlation between the time of white-light enhancement  and the peak time of hard X-ray (HXR) and microwave radiations. Also the Balmer lines are often very broad and strong (Fang \& Ding 1995). For type II WLFs, there are no such characteristics (Ding et al. 1999a, 1999b). This classification implies that WLFs may have different origins of WL emission and different heating mechanisms.

Temporal and spatial relationship between the enhancements of white light, HXR and radio emissions has been frequently investigated. WL kernels are usually found to be co-spatial with hard X-ray sources (Hudson et al. 1992; Metcalf et al. 2003; Chen \& Ding 2005, 2006; Krucker et al. 2011; Hao et al. 2012; Cheng et al. 2015; Kuhar et al. 2016; Yurchyshyn et al. 2017). Mart\'inez Oliveros et al. (2012) calculated the centroidal heights of the HXR source and WL emission for a flare close to the solar limb, and found that the mean heights above the photosphere are 305 $\pm$ 170 km and 195 $\pm$ 70 km, respectively. Krucker et al. (2015) studied three WLFs at the solar limb and found that the centroids of WL and HXR ($\ge30$ keV) sources share a similar height, which is about 300$-$450 km above the limb. Watanabe et al. (2010) studied an X1.5 WLF and found that the electron acceleration is closely correlated to the white-light production in time, space and power. Recently, Huang et al. (2016) studied 25 stronger flares  including 13 WLFs and found that the population of high energy electrons is larger when the WL emission is stronger. Kleint et al. (2016) investigated an X1 WLF and found that the energy deposited by electrons was sufficient for the extra ultraviolet (UV) and visible continuum emission. Lee et al (2017) analyzed an X1.6 WLF and suggested that the WL emmision enhancement was directly produced by non-thermal electrons. Based on a statistical analysis of 43 WLFs, Kuhar et al. (2016) found that the electrons of ~50 keV are the main energy source for WL emission. All these  reported WLFs are X- or M-class flares. 

Comparatively, enhanced WL emission has been less frequently observed in smaller flares. However, with the increasing sensitivity of detectors, more and more small WLFs have been discovered. Hudson et. al (2006) studied 11 WLFs including 4 C-class flares using observations obtained with the \emph{Transition Region and Coronal Explorer} (\emph{TRACE}; Handy et al. 1999) and found that the minimum enhancement of WL emission is about 8\%. Another example is given by Jess et al.(2008) who observed a C2.0 WLF. The duration of the WL emission is about 2 minutes and the diameter of the white-light kernel is less than $0.5^{\prime\prime}$. The enhancement of WL emission is above 300\%. In these cases, the WL sources occupy only a very small fraction of area of the entire flare ribbons. At present, there are many questions that are unclear and require further study, such as, how these small WLFs are produced, whether they belong to type-I or type-II WLFs, and what special conditions in the WL sources are required compared to the non-WL flare ribbons.

Although many more WLFs have been discovered in recent years than before,  it is still unknown how frequently WLFs occur. In previous studies, WLFs are casually captured by visual inspection from either images or spectra. Doing so inevitably misses some WLFs with weak intensities and small sizes. Therefore, to explore the above question, a systematic survey of WL emission for an active region with continuous observations of a few days is required.

Magnetic field changes associated with solar flares have been reported in many previous studies (e.g., Severny 1964; Tanaka 1978; Patterson 1984; Chen et al. 1989; Kosovichev \& Zharkova 2001). These changes are usually classified into two categories. One is flare-associated rapid and permanent changes which are thought to be real changes of magnetic field, due to the fact that the magnetic fields become more horizontal at the polarity inversion lines of flaring regions (Hudson et al. 2008; Wang \& Liu 2010; Fisher et al. 2012). The other is the so-called magnetic transients (MTs; e.g., Kosovichev \& Zharkova 2001; Qiu \& Gary 2003; Zhao et al. 2009), which are usually believed to be an observational artifact produced by the changes of the spectral line profiles during flares. MTs are generally found near the flare loop footpoints and persist only for a very short period of time.

In this paper, we investigate WLFs and their relationship with MTs in NOAA active region 11515 by using data taken by the \emph{Helioseismic and Magnetic Imager} (\emph{HMI}; Scherrer et al. 2012, Schou et al. 2012a, 2012b) and the \emph{Atmospheric Imaging Assembly} (\emph{AIA}; Lemen et al. 2012) on board the \emph{Solar Dynamics Observatory} (\emph{SDO}; Pesnell et al. 2012). We use HMI continuum intensity images to identify WLFs, and use the HMI line-of-sight magnetograms to detect the magnetic field changes. Our observations are presented in Section 2. Analysis and results are given in Section 3. In Section 4 we present a brief summary and discussion.

\section{Observations}

In this study, we use HMI full-disk continuum filtergrams and line-of-sight magnetograms, both observed by using the line of Fe {\footnotesize I} 6173~\AA\ with a 4096 $\times$ 4096 CCD detector, to identify WLFs and MTs. The Fe {\footnotesize I} 6173~\AA\  line is an optical line formed in the photosphere. The temporal cadence of the data is 45 s and the spatial resolution is about $1^{\prime\prime}$. AIA 131 \AA\ , 171 \AA\  and 1600 \AA\ images are used to show the morphology of the flares in the upper atmosphere e.g., Figure 4). The AIA 1600 \AA\ images are also used to determine the flare regions (see below). The temporal cadence of the AIA observations is 12 s in the extreme ultraviolet (EUV) passbands and 24 s in the ultraviolet (UV) passbands. The spatial resolution is about  $1.5^{\prime\prime}$.

It should be noted that the HMI continuum intensity ($I_{c}$) is obtained by ``reconstructing'' the spectral line through the following equation:
\begin{equation}
\emph{$I_{c}=\displaystyle\frac{1}{6}~\displaystyle{\sum_{j=0}^5}~[I_j+I_d~exp(-\frac{(\lambda-\lambda_0)^2}{\sigma^2})]$}\label{equation1},
\end{equation}
where $\lambda_0$, $\sigma$ and $I_d$ are the rest wavelength, line width and line depth, estimated by taking six sampling points ($I_j$) across the Fe {\footnotesize I} absorption line 6173.3 \AA\ (Couvidat et al., 2012).

NOAA active region 11515 was very active, producing nearly one hundred flares during its passage on the solar disk. However, there were no X-class flares. In this study we only consider flares that occurred between $E45^\circ$ and $W45^\circ$ to avoid possible problems with projection effects. In total 70 flares above C class were recorded during this period.
 
Usually it is easier to identify WL emission enhancement in flares with a high GOES class, whereas it is difficult to detect WL emission enhancement in low-class flares since normally the WL emission in such flares is only weakly enhanced above the quiescent values. To clearly show the small changes in the WL emission, we have constructed pseudo-intensity images by magnifying the difference between two adjacent continuum filtergrams by a factor of 5. The pseudo-intensity is expressed as: $I^\prime_{t0+45s}=(I_{t0+45s}-I_{t0})\times5+I_{t0}$, where $I_{t0}$ and $I_{t0+45s}$ are the original continuum intensity at two adjacent times with a gap of 45 s (the observing cadence of HMI). From the pseudo-intensity images, it is much easier to detect WLFs even when they are very weak (Figure 1). To demonstrate the capability of our method, online animations of a WLF using the original HMI continuum intensity images and using the pseudo intensity images, respectively, are provided. It should be noted that this method is only used to help determine whether there is an impulsive enhancement of WL emission during a flare. 

Through this approach, 20 out of the 70 flares are found to reveal an enhancement in the WL emission, though with different sizes of the enhancement area. Following the definition of WLFs from many recent investigations (e.g., Krucker et al. 2015; Huang et al. 2016;  Kuhar et al. 2016), we define these 20 flares which show an impulsive enhancement in HMI continuum intensity as WLFs. Twelve  of them are found in M-class flares and the other eight are found in C-class flares. Table 1 lists the observational information and results for these WLFs.

\begin{table*}[htbp]
\centering
\begin{threeparttable}
\centerline{\footnotesize Table 1. Information of the white-light flares detected in NOAA AR 11515}
\label{tab1}
{\tiny
\begin{tabular}{cccccccccccc}
\hline
\hline
Num &Date&Peak&GOES& AR &$\triangle T_{wl}$&$S_{wl}$&$dI^{m}_{wl}$&$dI^{a}_{wl}$&$dB_l$&$B_l$ sign&$I^{wl}_{1600}$\\
&&Time&Class& Location &(min)&(${1^{\prime\prime}}^2$)&$((I^{p}_{wl}-I^0_{wl})/I^0_{wl})$&$((I^{p}_{wl}-I^0_{wl})/I^0_{wl})$&$((B_l^{p}-B_l^0)/B_l^0)$&change&$(I_{1600}/I_{1600}^{max})$\\
\hline
1  &2012.07.03&17:02&C9.0&S17W08&3.75&10.50&$9.3\%\pm1.1\%$&$6.5\%\pm1.1\%$&$1.4\%$&Y&$72.2\%$\\
2  &2012.07.04&09:55&M5.3&S17W18&5.25&38.00&$32.6\%\pm1.2\%$&$8.8\%\pm1.2\%$&$1.7\%$&	Y &$82.5\%$\\
3  &2012.07.04&12:24&M2.3&S17W22&3.00&3.75&$8.0\%\pm1.3\%$&$5.7\%\pm1.3\%$&$5.7\%$	&Y	&$85.2\%$\\
4  &2012.07.04&14:40&M1.3&S18W20&4.50&32.25&$16.1\%\pm1.0\%$&$7.5\%\pm1.1\%$&$-6.0\%$	&Y	&$66.0\%$\\
5  &2012.07.04&15:50&C6.4&S17W23&3.00&13.75&$9.0\%\pm1.3\%$&$6.4\%\pm1.3\%$&$-7.0\%$	&N	&$84.5\%$\\
6  &2012.07.04&16:12&C6.9&S17W21&3.75&35.00&$13.3\%\pm1.2\%$&$7.1\%\pm1.2\%$&$-1.7\%$	&N	&$50.8\%$\\
7  &2012.07.04&21:27&C9.5&S16W21&3.00&34.50&$16.5\%\pm1.3\%$&$9.0\%\pm1.3\%$&$-16.2\%$   &Y	&$80.3\%$\\
8 &2012.07.04&22:09&M4.6&S16W28&3.75&14.25&$13.8\%\pm1.1\%$&$7.5\%\pm1.1\%$&$-5.1\%$	&---	&$93.2\%$\\
9 &2012.07.05&01:10&M2.4&S17W27&4.50&26.00&$10.1\%\pm1.4\%$&$6.7\%\pm1.4\%$&$-0.7\%$	&---	&$64.6\%$\\
10&2012.07.05&02:42&M2.2&S18W27&5.25&35.00&$11.2\%\pm1.3\%$&$6.7\%\pm1.3\%$&$-0.4\%$	&Y	&$74.4\%$\\
11&2012.07.05&03:36&M4.7&S18W29&4.50&68.75&$49.9\%\pm1.1\%$&$14.8\%\pm1.1\%$&$-30.2\%$&	Y&$98.0\%$\\
12&2012.07.05&04:45&C9.1&S18W29&2.25&38.00&$14.0\%\pm1.2\%$&$6.9\%\pm1.2\%$&$-1.7\%$	&---	&$60.7\%$\\
13&2012.07.05&10:48&M1.8&S18W30&5.25&37.25&$27.0\%\pm1.3\%$&$10.9\%\pm1.3\%$&$-12.0\%$&	Y&	$91.5\%$\\
14&2012.07.05&11:44&M6.1&S18W32&6.00&165.0&$36.0\%\pm1.1\%$&$10.3\%\pm1.1\%$&$-10.1\%$&	Y&	$93.0\%$\\
15&2012.07.05&14:45&C8.3&S17W33&5.25&4.25&$9.6\%\pm1.1\%$&$7.5\%\pm1.1\%$&$-13.1\%$	&Y	&$91.9\%$\\
16&2012.07.05&15:59&C6.2&S17W34&5.25&19.25&$18.7\%\pm1.3\%$&$8.3\%\pm1.3\%$&$0.3\%$	&Y	&$90.2\%$\\
17&2012.07.05&20:14&M1.6&S18W37&6.00&30.00&$14.0\%\pm1.2\%$&$7.7\%\pm1.2\%$&$-20.3\%$	&Y	&$94.2\%$\\
18&2012.07.06&01:40&M2.9&S17W39&6.00&94.25&$40.2\%\pm1.1\%$&$10.0\%\pm1.1\%$&$13.5\%$	&Y	&$81.7\%$\\
19&2012.07.06&07:07&C7.4&S18W44&6.00&20.75&$12.1\%\pm1.1\%$&$7.2\%\pm1.1\%$&$-1.2\%$&---	&$78.0\%$\\
20&2012.07.06&10:32&M1.8&S17W44&6.75&23.25&$11.2\%\pm1.1\%$&$6.9\%\pm1.1\%$&$-4.2\%$	&---	&$66.2\%$\\

\hline
\end{tabular}
}
\begin{tablenotes}
        \footnotesize
        \item[1] $\triangle T_{wl}$ is the duration of the WLF. $S_{wl}$ is the area of the WLF region.
        \item[2] $I^0_{wl}$ and $B_l^0$ are the intensity and line-of-sight magnetic field strength before the flare peak in the WLF region. $I^{p}_{wl}$ and $B_l^{p}$ are the intensity and line-of-sight magnetic field strength at the peak time of the flare in the WLF region. $dI_{wl}^{m}$ refers to the maximum value of $(I^{p}_{wl}-I^0_{wl})/I^0_{wl}$ in the WLF region and $dI_{wl}^{a}$ refers to the average value. $dB_l$ is the average value of $(B_l^{p}-B_l^0)/B_l^0$ in the WLF region.
        \item[3] For the sign change of $B_l$, `Y' means clear sign change, `N' means no sign change, and `---' means not obvious.
        \item[4] $I_{1600}$ is the average intensity of AIA 1600\AA\ in the WLF region and $I^{max}_{1600}$ is the maximum value in the whole flare region.
      \end{tablenotes}
\end{threeparttable}
\end{table*}

\section{Analysis and results}

The dates, peak times, GOES classes and positions of these 20 WLFs can be seen from Table 1. For each WLF, we first define the WLF region. To estimate the fluctuation of the background, we select three quiet-sun regions (R1, R2, R3; marked in Figure 2(b) \& (d)) outside the sunspots, which are located in the north, south and east of the active region, respectively. These three regions are nearly devoid of field during the periods when we estimate the fluctuation. All three regions have a size of $20\times20$ pixels, which would cover several granules. We then calculate the standard deviations of the running difference within each of these three regions during three different periods. From Figure 3 we can see that the standard deviations in these three regions during different periods are very close. The average value is about 0.013 and the maximum value is about 0.016, which can be regarded as a level of the fluctuation of the background. We define the WLF region as the area where the emission enhancement ($(I^{p}_{wl}-I^{0}_{wl})/I^0_{wl}$) is greater than 0.05 ($>3\times0.016$). The average intensity of the WL emission in the WLF region at each time is also calculated to obtain a light curve of the WL emission. The emission curve usually shows a pulse-like structure (see Figure 6). We define the start and end of the pulse as the start and end times of the WLF, which gives the duration of WLF ($\triangle T_{wl}$). We calculate $dI^a_{wl}$, which is the average value of the percentage of WL intensity increase at the peak of the flare relative to that before the peak of flare. The maximum value of this percentage, $dI^m_{wl}$, is also calculated. The error of WL enhancement is estimated as the average of the standard deviations of the changes of WL emission in the three quiet-sun regions (R1, R2, R3) at the same time, and it reflects the intensity fluctuation of the granules. Similar to $dI^a_{wl}$, $dB_l$ is the average value of the percentage change of unsigned line-of-sight magnetic field in the WLF region. Also, we calculate the ratio between the average intensity of the WLF region in the AIA 1600 \AA\ images, and the maximum intensity in the 1600 \AA\ images within the whole flaring region during the peak time of the flare ($I^{wl}_{1600}$). Using this ratio, we can determine whether the WLF occurred in the central area of the flare or elsewhere. The central area of a flare is defined as the region where the AIA 1600 \AA\ intensity is greater than half the maximum value at the peak time. It should be noted that the central  area of a flare refers to the central area of the whole flaring region.

From Table 1 we can see that these WLFs generally have a short duration, with an average lifetime of 4.65 minutes. The size of the observed WLFs are generally small, with an average area of about 37.18 square arcseconds. From the percentage increase of the continuum intensity ($dI^a_{wl}$, $dI^m_{wl}$), we can see that the WL enhancements in these WLFs are generally very weak. Hudson et al. (2006) detected 11 WLFs in the WL channel of TRACE, including 4 C-class flares. The minimum excess contrast is only 0.08 $\pm$ 0.017 for the C1.6 WLF on 2004 July 24. But for all the other 10 WLFs the excess contrast exceeds 0.10. Using observations of \emph{RHESSI} and HMI, Kuhar et al. (2016) studied 43 WLFs (M- \& X- classes) and found that the lowest change of white-light emission is 0.08 $\pm$ 0.07. It should be noted that the change of WL emission in these papers refers to the brightest pixel in the images of relative enhancement. This parameter, denoted as $dI^m_{wl}$ in Table 1, is greater than $20\%$ for only 5 WLFs in our study. The lowest value of $dI^m_{wl}$ is $8\%\pm1.3\%$, which is comparable to the lowest values given by Hudson et al. (2006) and Kuhar et al. (2016). However, the average changes of the WL emission in WLF regions ($dI^a_{wl}$) are mostly less than 10\%, with an average of 8.1\% and a minimal of $5.7\%\pm1.3\%$. Only 3 WLFs reveal an average intensity enhancement greater than 10\%, and the largest is $14.8\%\pm1.1\%$. The values of $I_{1600}^{wl}$ for all WLFs are greater than 0.5, meaning that the WL enhancements all occur in the central areas of the flare ribbons defined by the enhanced 1600 \AA\ emission.

Figure 2 shows the occurrence times and locations of these WLFs. In panel (a) the solid line is the soft X-ray (1-8\AA) flux measured by GOES. The dotted vertical lines mark the peak times of the 70 flares detected in  AR 11515 between $E45^\circ \sim W45^\circ$ on the solar disk, where the red and black ones are for the WLFs and normal flares, respectively. Panels (b) and (d) are the continuum intensity images at two different times. Panels (c) and (e) are the corresponding line-of-sight magnetgrams. In panels (b)-(e) the red circles mark the central positions of these WLFs. From panel (a) and Table 1 we can see that these WLFs occur mainly on the days of July 4th and July 5th. From panels (b)-(e) we can see a narrow ribbon-like magnetic field structure in active region 11515 on these two days. The magnetic field at the ribbon is negtive, but the magnetic field on both sides of the ribbon are positive. All the WLFs are distributed along this ribbon.

Figure 4 shows the white-light difference images for these 20 WLFs. The red contours in each panel mark the WLF region we defined based on the method described above.  All these images have a size of $ \sim40^{\prime\prime}\times40^{\prime\prime}$, corresponding to the main flaring region and covering the whole WL enhancement regions. It should be noted that the WL difference images are obtained by calculating $(I^p_{wl}-I^0_{wl})/I^0_{wl}$, where $I^p_{wl}$ is the HMI continuum intensity at the peak time of the flare and $I^0_{wl}$ is the intensity several minutes before the flare peak.  

Figure 5 shows an individual example for an M4.7 WLF which occurred on July 5th. Panels (a)-(c) show the AIA 1600\AA, 171\AA\ and 131\AA\ images at the peak time of this WLF, respectively. From these three panels we can see that the spatial scale of this flare is small. The white box in panel (a) corresponds to the field of view (FOV) in panels (d)-(i). Panels (d) and (e) are the continuum images at the time before and at the peak time of the flare. Panel (f) is the difference image between them. Panels (g)-(i) are similar to panel (d)-(f) but for the magnetic field. The green box in panels (d)-(i) shows the region where the WLF and MT occurred. From these panels we can see that there is a significant enhancement of the continuum intensity in a very short time period and the HMI line-of-sight magnetic field changes significantly at the same time and the same locations. In other words, WLF and MT occurred simultaneously and co-spatially. Note that the changes of both the continuum intensity and magnetic field are transient and not permanent.

Figure 6 shows the temporal evolution of the white light intensity (red) and unsigned line-of-sight magnetic field (blue) around the flaring times of these 20 WLFs. Each diamond on these curves represents the average value of continuum intensity or unsigned magnetic field in the WLF region. The green shaded region in each panel marks the duration of the WLF, which is defined as the period between times when the WL enhancement obviously appeared and vanished. To estimate the uncertainty of the WL intensity, we first plot the WL curves in three quiet-sun regions (R1, R2, R3) during the same period of the flare. Then we take the average value of the three standard deviations derived from the three WL curves as the intensity uncertainty for the WLF. The figure clearly shows that the MTs and WLFs occur simultaneously. From Table 1 we can see that MTs in 13 WLFs show an obvious change of sign. It should be noted that the MT in the WLF 12 (the C9.1 WLF on Jul 5th) seems to be very weak and difficult to be identified. This is because the transient mainly occurred in regions of negative field, the strength of which is much weaker compared to that of the positive field. Thus, the MT is not reflected obviously in the curve of unsigned magnetic field for this WLF. Also, we find a close relationship between the WL enhancement and magnetic field change, which is consistent with the conclusion of Song \& Zhang (2016) that the magnetic field and intensity variations are closely related and they are possibly the two facets of the same phenomena of a solar flare. It should be noted that the WL enhancement and magnetic field change in their observations are permanent. MTs are usually believed to be artifacts due to the distortions of line profile (Patterson 1984, Ding et al. 2002, Qiu  \& Gary  2003, Isobe et al. 2007, Mauraya et al. 2012), although some recent studies suggest that some of them are real (Matthews et al. 2011, Harker \& Pevtsov 2013). In this study, we do not investigate whether MTs are real or not, which relies on a detailed future examination of the profile of the 6173~\AA\ line.

In Figure 7, panel (a) shows the relationship between the change of WL emission ($dI^a_{wl}$) and the AIA 1600\AA\ intensity ($I_{1600}$) in WLF regions, while panel (b) shows the relationship between the change of WL emission ($dI^a_{wl}$) and the absolute change of unsigned line-of-sight magnetic field ($|dB_l|$). In panel (a), we can see that the values of $I_{1600}^{wl}$ in all WLFs are greater than 0.5, which means that all of these WLFs occur at the central regions of the flare ribbons. There is a trend that the location of the WLF is closer to the center of the whole flaring region when the enhancement of the WL emission is stronger. Panel (b) reveals a possible linear correlation between the changes of WL emission and unsigned magnetic field. A greater enhancement of the WL emission is often accompanied by a larger change of the line-of-sight magnetic field. This can be seen from Table 1. 


\section{Summary and discussion}

WLFs were believed to be very rare compared to the frequent occurrence of solar flares. However, in this study we find at least 20 WLFs out of the 70 flares in NOAA AR 11515 during its passage on the solar disk ($E 45^\circ \sim W45^\circ$). Thus, the occurrence rate of WLFs in this active region is at least 28.6\%, which provides further evidence towards the idea that all flares may have a WL component, a possibility discussed by Hudson et al. (2006) several years ago. Our findings also emphasize the fact that WLFs can occur in less energetic flares, i.e, C- and lower M-class flares, following some previous studies (e.g., Matthews et al. 2003; Hudson et al. 2006; Jess et al. 2008; Kowalski et al. 2015a). For most of these 20 WLFs, the WL enhancement lasts for a short duration and occurs in a small region. The average enhancement ($dI^a_{wl}$) of the WL emission in the WLF region is generally very small, with an average of 8.1\%. If we regard a WLF whose change of WL emission ($dI^a_{wl}$) is lower than 10\% as a weak one, then there are 17 weak WLFs in our sample and the percentage is 85\% (17/20). This suggests that most WLFs are very weak, which may explain why WLFs are not frequently reported. Our results appear to support the argument given by Jess et al. 2008 that the sensitivity and resolution of a telescope will limit the detection of the WL emission.

All the WLFs appear in the central areas ($I_{1600}^{wl}>0.5$) of the flare ribbons. This can be understood as the central  areas of flare ribbons usually have a greater energy deposition and WL enhancement most likely occurs there. All 20 WLFs in NOAA AR 11515 occur mainly on 4 July and 5 July 2012, and are distributed along a narrow ribbon with negative magnetic flux surrounded by positive flux on both sides. The high occurrence rate of WLFs in this active region may be related to the development of this special magnetic field configuration. Our results appear to support the conclusion of Neidig \& Cliver (1983) that WLFs are often produced in large and magnetically complex active regions. 

Another interesting result is that these 20 WLFs are all accompanied by MTs. For the remaining 50 flares without obvious WL enhancement, there are no detectable MTs. It is still an open question whether MTs are real or not (see Harker \& Pevtsov (2013) and references therein). Nevertheless, our observations suggest that MTs and WLFs are closely related when we consider the corresponding HMI data for this particular active region. They occur at the same time and at the same location. A larger enhancement of WL emission is often accompanied by a lager change of the line-of-sight component of the unsigned magnetic field. Considering the close relationship between MTs and WLFs, many previously reported flares with MTs may be WLFs.

Although many more WLFs are detected in AR 11515, they are mostly small-scale and short-lived brightenings in the 6173~\AA\ continuum, similar to the small WL kernel reported by Jess et al. (2008). In the future we plan to examine more ARs to see if most WLFs possess similar features. Also, it should be noted that the HMI continuum intensity ($I_{c}$) is obtained by ``reconstructing'' a spectral line. Although the enhancement in the HMI continuum intensity is commonly identified as a signature of WLFs (e.g., Krucker et al. 2015; Huang et al. 2016;  Kuhar et al. 2016), we cannot exclude the possibility that the HMI $I_c$ value may have been, to a certain degree, affected by the emission in the line core. 

In addition, several questions need to be clarified with future observations. First, what is the energy source of such small-scale and short-lived WLFs? If they are powered by electrons beams, as in the type-I WLFs studied previously, they should accompany small bursts in the hard X-ray (or microwave) emission. If the heating energy is from other sources like Alfv\'en waves (Fletcher \& Hudson 2008) or even in situ energy release, there may not be obvious correlation between WL enhancement  and the hard X-ray emission. The latter case belongs to type-II WLFs. Simultaneous observations of spectral lines formed at different layers are helpful in judging the energy source. To answer this question coordinated observations of several instruments with very high cadence and spatial resolution are required. Second, what is the relationship between such a small WL kernel and the whole flare? What special condition is required to produce a WLF? For this purpose, vector magnetic field observations and three-dimensional magnetic field extrapolation are required to check if the WLFs are located at peculiar sites where either magnetic reconnection likely occurs or energy can be easily transported there.


\begin{acknowledgements}
This work is supported by NSFC grants 11790304(11790300), 11125314 and 11733003, the Recruitment Program of Global Experts of China, the Specialized Research Fund for State Key Laboratories and the Max Planck Partner Group program. H.T. acknowledges support of ISSI and ISSI-BJ to the team ``Diagnosing heating mechanisms in solar flares through spectroscopic observations of flare ribbons''. We thank the anonymous reviewer for the carefully reading and very constructive comments. 
\end{acknowledgements}

\newpage

\begin{figure*}[!ht]
\centerline{\includegraphics[width=0.95\textwidth]{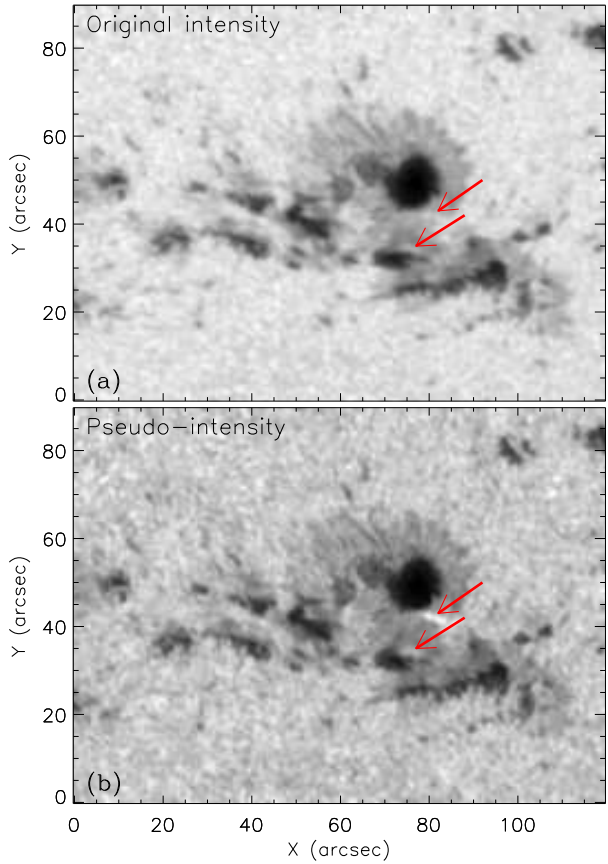}}
\caption{HMI continuum intensity images of an M5.3 WLF observed at 09:54:53 UT on 2012 July 4. (a) is the original intensity and (b) shows the pseudo-intensity. Red arrows point out the regions where the WL enhancement occurred. (An animation of this figure is available.)}
\end{figure*}

\begin{figure*}[!ht]
\centerline{\includegraphics[width=0.9\textwidth]{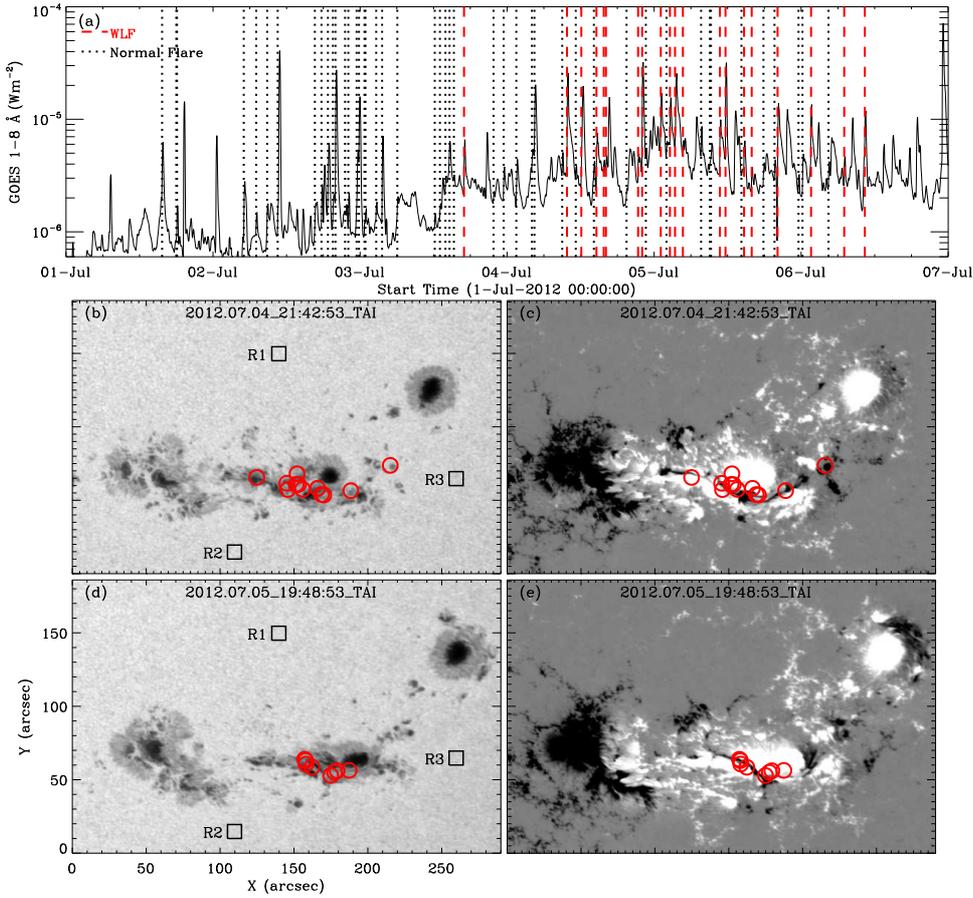}}
\caption{White-light flares (WLFs) in NOAA AR 11515. (\emph a) 70 flares (C- and M- classes) occurred between $E45^\circ$ to $W45^\circ$. The solid line is GOES soft X-ray (1-8 \AA). The red dashed lines mark the peak times of WLFs, and the black dotted lines mark these of normal flares. (\emph b)-(\emph e)  Locations of these WLFs in the active region. The circles in panels (b) and (c) show WLFs 1--12 while the circles in panels (d) and (e) show WLFs 13--20. (\emph b) and (\emph d): HMI continuum images at two different times, (\emph c) and (\emph e): HMI line-of-sight magnetgrams at two different times. The black boxes R1, R2 and R3 in panels (b) and (d) mark the quiet-sun regions selected to estimate the error in the measured continuum intensity. The size of each box is $20\times20$ pixels.}
\end{figure*}

\begin{figure*}[!ht]
\centerline{\includegraphics[width=0.45\textwidth]{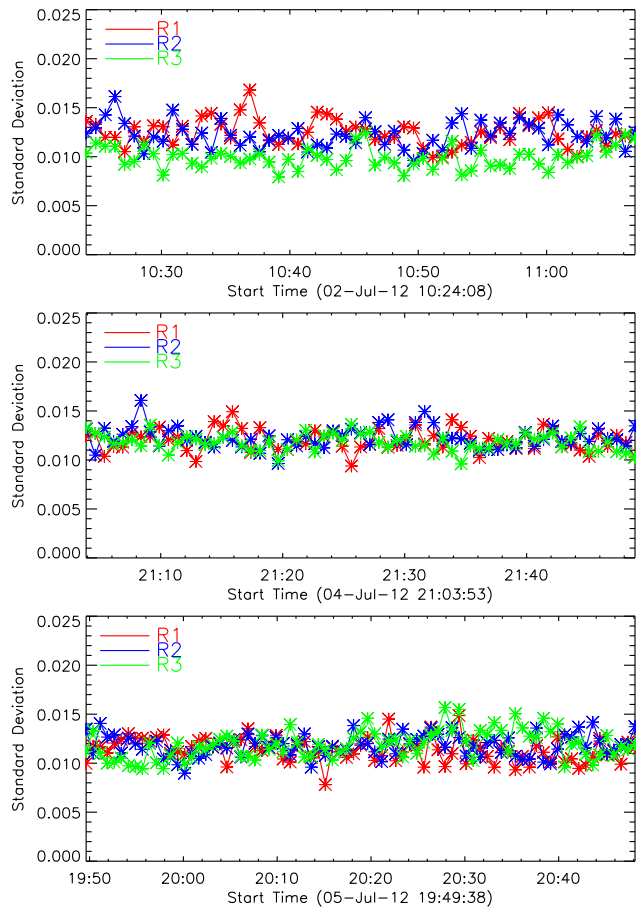}}
\caption{Standard deviations of the difference of the HMI continuum intensity in three quiet-sun regions (R1, R2 and R3) during three different periods (top, middle, bottom). The red, blue and green colors represent the results in the regions of R1, R2 and R3, respectively. }
\end{figure*}

\begin{figure*}[!ht]
\centerline{\includegraphics[width=0.9\textwidth]{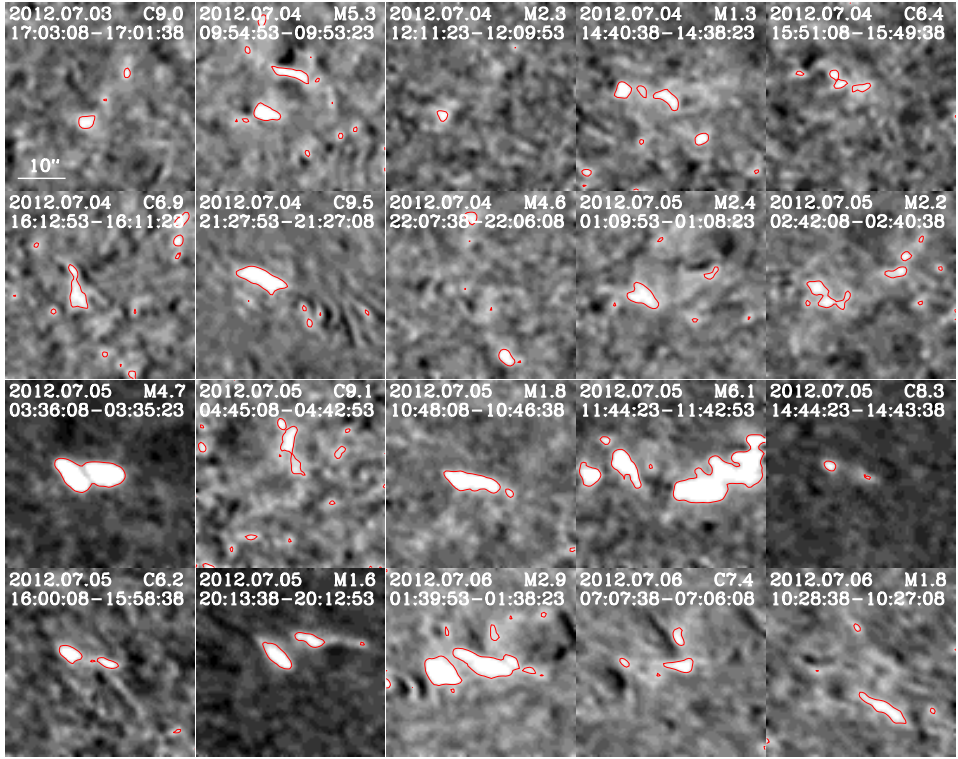}}
\caption{ Snapshots of 20 WLFs. The images are obtained by calculating $(I^p_{wl}-I^0_{wl})/I^0_{wl}$, where $I^p_{wl}$ and $I^0_{wl}$ are the intensity images taken at two different times. The red contours mark the WLF regions. Each image has a size of $ \sim40^{\prime\prime}\times40^{\prime\prime}$.} 
\end{figure*}

\begin{figure*}[!ht]
\centerline{\includegraphics[width=0.9\textwidth]{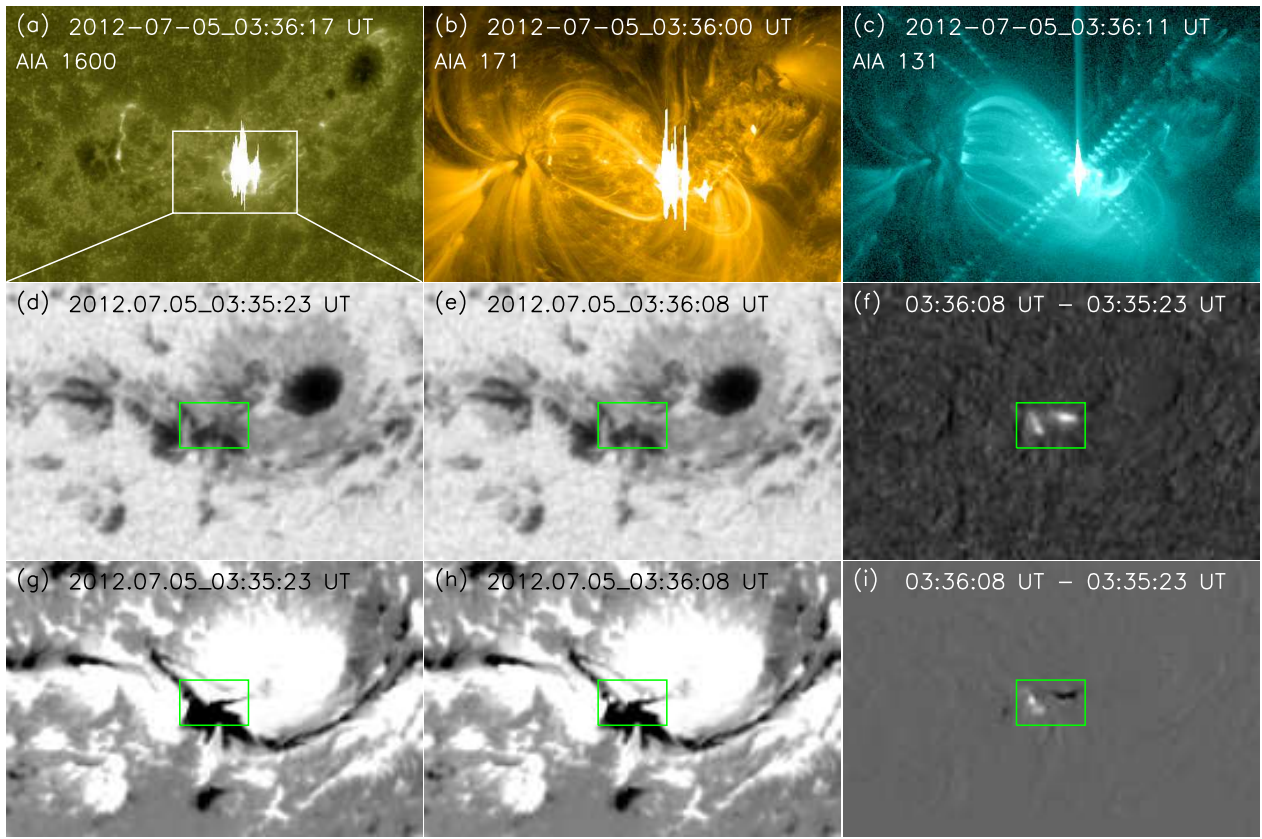}}
\caption{A white-light flare in NOAA AR 11515.  (\emph a), (\emph b) and (\emph c) are AIA 1600, 171 and 131 \AA\ images at the peak time of this WLF, respectively. (\emph d) and (\emph e) are HMI continuum images at the beginning and peak time of this WLF, respectively. (\emph f) is the difference image between (\emph d) and (\emph e). (g) and (h) are HMI line-of-sight magnetgrams at the beginning and peak time of this WLF, respectively. (\emph i) is the difference image between (\emph g) and (\emph h). The white box in panel (a) corresponds to the FOV in panels (d)-(i). The green box in panels (\emph d)-(\emph i) marks the location where the flare occurred. }
\end{figure*}

\begin{figure*}[!ht]
\centerline{\includegraphics[width=0.9\textwidth]{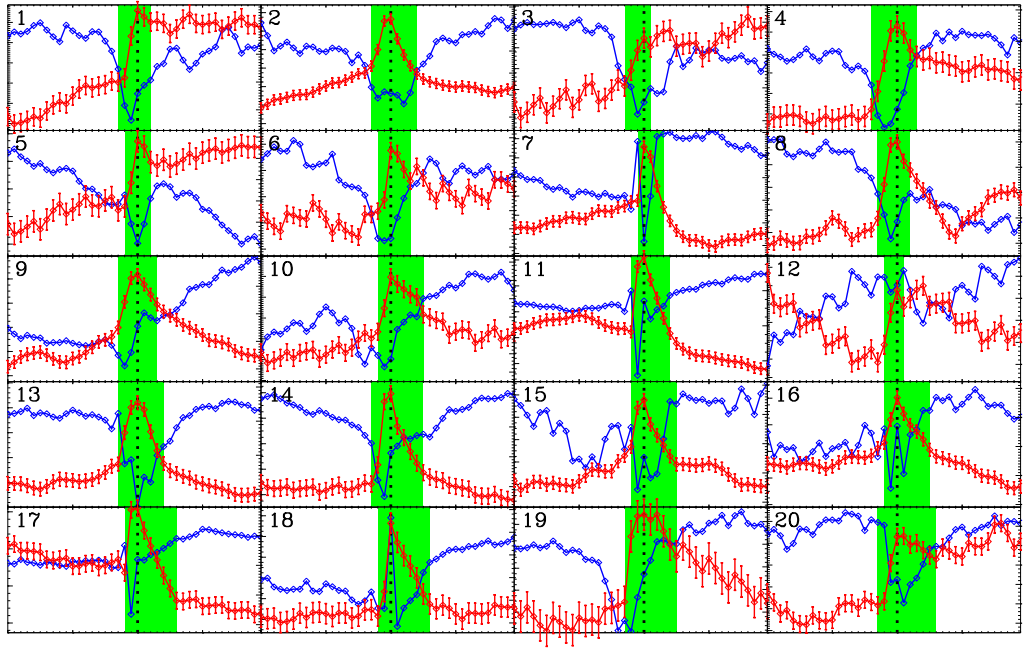}}
\caption{Temporal evolution of the HMI continuum (red) and unsigned $B_l$ (line-of-sight magnetic field, blue) for the 20 WLFs. The black dotted line marks the time of the WL emission peak for each WLF. The green shaded region marks the duration of each WLF. The time range shown in each panel is from 25 minutes before the peak time to 25 minutes after it.  }
\end{figure*}

\begin{figure*}[!ht]
\centerline{\includegraphics[width=0.45\textwidth]{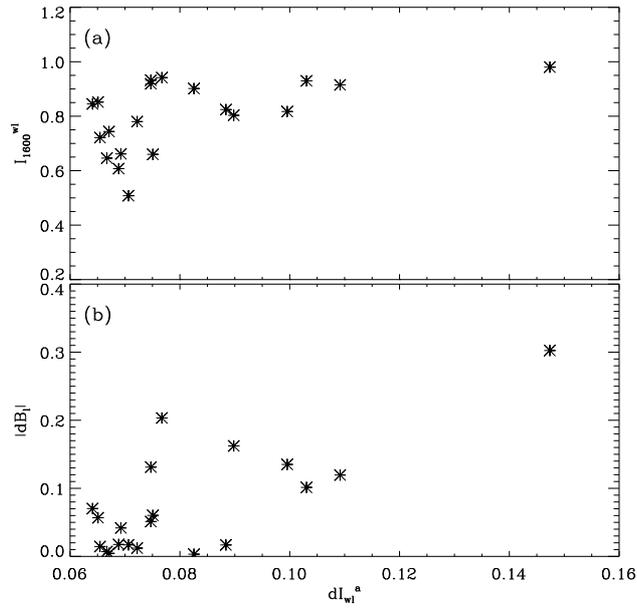}}
\caption{(a) The relationship between WL enhancement ($dI_{wl}^a$, the average enhancement of WL emission in the WLF region) and AIA 1600~\AA\ intensity ($I_{1600}^{wl}$, the ratio between the average AIA 1600~\AA\  intensity in the WLF region and the maximum 1600~\AA\  intensity in the whole flare region) for 20 WLFs in NOAA AR 11515. The positive correlation indicates that a WLF has a stronger WL emission if it occurs closer to the center of the whole flaring region. (b) The relationship between WL enhancement ($dI_{wl}^a$)  and magnetic field change ($|dB_l|$, the absolute change of the unsigned $B_l$ in WLF region). A greater enhancement of the WL emission is often accompanied by a larger change of the line-of-sight magnetic field.}
\end{figure*}


\begin{thebibliography}{}

\bibitem[Aboudarham et al.(1986)]{Aboudarham1986}
Aboudarham, J., \& H\'enoux, J.C. 1986, A\&A, 168, 301

\bibitem[Carrington (1859)]{Carrington1859}
Carrington, R. C. 1859, MNRAS, 20, 13

\bibitem[Chen et al.(2005)]{Chen2005}
Chen, Q. R., \& Ding, M. D. 2005, ApJ, 618, 537

\bibitem[Chen et al.(1989)]{Chen1989}
Chen, J., Ai, G., Zhang, H., et al. 1989, PYunO, 1, 108

\bibitem[Cheng et al.(2015)]{Cheng2015}
Cheng, X., Hao, Q., Ding, M. D., et al. 2015, ApJ, 809, 46

\bibitem[Couvidat et al.(2012)]{Couvidat2012}
Couvidat, S., Rajaguru, S.P., Wachter, R., et al, 2012, SoPh, 278, 217

\bibitem[Ding et al.(1999a)]{Ding1999a}
Ding, M. D., Fang, C., Yin, S. Y., \& Chen, P. F. 1999a, A\&A, 348, L29

\bibitem[Ding et al.(1999b)]{Ding1999b}
Ding, M. D., Fang, C., \& Yun, H. S. 1999, ApJ, 512, 454

\bibitem[Emslie et al.(1982)]{Emslie1982}
Emslie, A. G., Sturrock, P. A., 1982, SoPh, 80, 99

\bibitem[Fang et al.(1995)]{Fang1995}
Fang, C., \& Ding, M. D. 1995, A\&AS, 110, 99

\bibitem[Fang et al.(2013)]{Fang2013}
Fang, C., Chen, P. F., Li, Z., et al. 2013, RAA, 13, 12

\bibitem[Fisher et al.(2012)]{Fisher2012}
Fisher, G. H., Bercik, D. J., Welsch, B. T., \& Hudson, H. S. 2012, SoPh, 277, 59

\bibitem[Fletcher et al.(2008)]{Fletcher2008}
Fletcher, L., \& Hudson, H. S. 2008, ApJ, 675, 1645

\bibitem[Gan et al.(1994)]{Gan1994}
Gan, W. Q. \& Mauas, P. J. D. 1994, ApJ, 430, 891

\bibitem[Harker et al.(2013)]{Harker 2013}
Harker, B. J., \& Pevtsov, A. A. 2013, ApJ, 778, 175

\bibitem[Handy et al.(1999)]{Handy1999}
Handy, B. N., Acton, L. W., Kankelborg, C. C. et al. 1999, SoPh. 187. 229

\bibitem[Hao et al.(2012)]{Hao2012}
Hao, Q., Guo, Y., Dai, Y., et al. 2012, A\&A, 544, L17

\bibitem[Henoux et al.(1977)]{ Henoux1977}
 H\'enoux, J. C., \& Nakagawa, Y. 1977, SoPh, 53, 219

\bibitem[Heinzel et al.(2014)]{ Heinzel2014}
Heinzel, P., \& Kleint, L. 2014, ApJL, 794, L23

\bibitem[Hodgson et al.(1859)]{Hodgson1859}
Hodgson, R. 1859, MNRAS, 20, 15

\bibitem[Huang et al.(2016)]{Huang 2016}
Huang, N. Y., Xu, Y., \& Wang, H. 2016, RAA, 16, 177

\bibitem[Hudson et al.(1972)]{Hudson1972}
Hudson, H. S., Ohki, K., 1972, SoPh, 23, 155.

\bibitem[Hudson et al.(1992)]{Hudson1992}
Hudson, H. S., Acton, L. W., Hirayama, T., \& Uchida, Y. 1992, PASJ, 44, L77

\bibitem[Hudson et al.(2006)]{Hudson2006}
Hudson, H. S., Wolfson, C.J., Metcalf, T.R. 2006, SoPh, 234, 79

\bibitem[Hudson et al.(2008)]{Hudson2008}
Hudson, H. S., Fisher, G. H., \& Welsch, B. T. 2008, in ASP Conf. Ser. 383, Subsurface and Atmospheric Influences on Solar Activity, ed. R. Howe et al. (San Francisco, CA: ASP), 221

\bibitem[Isobe et al.(2007)]{Isobe2007}
Isobe, H., Kubo, M., Minoshima, T., et al. 2007, PASJ, 59, 807

\bibitem[Jess et al.(2008)]{Jess2008}
Jess, D. B., Mathioudakis, M., Crockett, P. J., Keenan, F. P., 2008, ApJL, 688, L119

\bibitem[Kleint et al.(2016)]{Kleint2016}
Kleint, L., Heinzel, P., Judge, P., \& Krucker, S. 2016, ApJ, 816, 88

\bibitem[Kosovichev et al.(2001)]{Kosovichev 2001}
Kosovichev, A. G., \& Zharkova, V. V. 2001, ApJL, 550, L105

\bibitem[Kowalski  et al.(2015a)]{Kowalski  2015a}
Kowalski, A. F., Cauzzi, G., \& Fletcher, L. 2015, ApJ, 798, 107

\bibitem[Kowalski  et al.(2015b)]{Kowalski  2015b}
Kowalski, A. F., Hawley, S. L., Carlsson, M., et al. 2015, SoPh, 290, 3487

\bibitem[Krucker et al.(2011)]{Krucker2011}
Krucker, S., Hudson, H. S., Jeffrey, N. L. S., et al. 2011, ApJ, 739, 96

\bibitem[Krucker et al.(2015)]{Krucker2015}
Krucker, S., Saint-Hilaire, P., Hudson, H.S., et al. 2015, ApJ 802, 19.

\bibitem[Kuhar et al.(2016)]{Kuhar2016}
Kuhar, M., Krucker, S., Mart\'inez Oliveros, J.C. et al. 2016, ApJ, 816, 6

\bibitem[Lee et al.(2017)]{Lee2017}
Lee, K. -S., TImada, S., Watanabe, K. et al. 2017, ApJ, 836, 150

\bibitem[Lemen et al.(2012)]{Lemen2012}
Lemen, J. R., Title, A. M., Akin, D. J. et al. 2012, SoPh, 275, 17

\bibitem[Machado et al.(1986)]{Machado1986}
Machado, M. E., Avrett, E. H., Falciani, R., et al. 1986, in The lower atmosphere of solar flares, ed. D. F. Neidig, 483

\bibitem[Machado et al.(1989)]{Machado1989}
Machado, M. E., Emslie, A. G., \& Avrett, E. H. 1989, SoPh, 124, 303

\bibitem[Martinez et al.(2012)]{Martinez2012}
Mart\'inez Oliveros, J.C., Hudson, H.S., Hurford, G.J. et al. 2012, ApJL. 753, L26

\bibitem[Matthews et al.(2003)]{Matthews2003}
Matthews, S. A., van Driel-Gesztelyi, L., Hudson, H. S., \& Nitta, N. V. 2003, A\&A, 409, 1107

\bibitem[Matthews et al.(2011)]{Matthews2011}
Matthews, S. A., Zharkov, S., \& Zharkova, V. V. 2011, ApJ, 739, 71

\bibitem[Maurya et al.(2012)]{Maurya2012}
Maurya, R. A., Vemareddy, P., \& Ambastha, A. 2012, ApJ, 747, 134

\bibitem[Metcalf et al.(1990)]{Metcalf1990}
Metcalf, T. R., Canfield, R. C., Avrett, E. H., \& Metcalf, F. T. 1990, ApJ, 350, 463

\bibitem[Metcalf et al.(2003)]{Metcalf2003}
Metcalf, T. R., Alexander, D., Hudson, H. S., \& Longcope, D. W. 2003, ApJ, 595, 483

\bibitem[Neidig et al.(1983)]{Neidig1983b}
Neidig, D. F. \& Cliver, E. W., 1983, SoPh, 88, 275

\bibitem[Neidig (1983)]{Neidig1983}
Neidig, D. F., \& Cliver, E. W. 1983, Air Force Geophysics Lab. Technical Report AFGL-TR-83-0257

\bibitem[Neidig et al.(1989)]{Neidig1989a}
Neidig, D. F. 1989, SoPh, 121, 261

\bibitem[Patterson et al.(1984)]{NPatterson1984}
Patterson, A. 1984, ApJ, 280, 884

\bibitem[Pesnell et al.(2012)]{Pesnell2012}
Pesnell, W. D.,  Thompson, B. J. \& Chamberlin, P. C., 2012, SoPh. 275, 3

\bibitem[Qiu et al.(2003)]{Qiu2003}
Qiu, J., \& Gary, D. E. 2003, ApJ, 599, 615

\bibitem[Scherrer et al.(2012)]{Scherrer2012}
Scherrer, P. H., Schou, J., Bush, R. I., et al. 2012, SoPh, 275, 207

\bibitem[Schou et al.(2012)]{Schou2012a}
Schou, J., Borrero, J. M., Norton, A. A., et al. 2012a, SoPh, 275, 327

\bibitem[Schou et al.(2012)]{Schou2012b}
Schou, J., Scherrer, P. H., Bush, R. I., et al. 2012b, SoPh, 275, 229

\bibitem[Severny(1964)]{Severny1964}
Severny, A. B. 1964, ARA\&A, 2, 363

\bibitem[Svestka(1966)]{Svestka1966}
\v{S}vestka Z., 1966, Space Sci. Rev., 5, 388

\bibitem[Svestka(1970)]{Svestka1970}
\v{S}vestka Z., 1970, SoPh. 13, 471.

\bibitem[Song (2016)]{Song2016}
Song, Y. L. \& Zhang, M. 2016, ApJ, 826, 173.

\bibitem[Tanaka(1978)]{Tanaka1978}
Tanaka, K. 1978, SoPh, 58, 149

\bibitem[Yurchyshyn et al.(2017)]{Yurchyshyn2017}
Yurchyshyn, V., Kumar, P., Abramenko, V., et al. 2017, ApJ, 838, 32

\bibitem[Wang (2010)]{Wang(2010)}
Wang, H., \& Liu, C. 2010, ApJ, 716, L195

\bibitem[Watanabe et al.(2010)]{Watanabe2010}
Watanabe, K., Krucker, S., Hudson, H., et al. 2010, ApJ, 715, 651

\bibitem[Zhao et al.(2009)]{Zhao2009}
Zhao, M., Wang, J. X., Matthews, S., et al. 2009, RAA, 9, 812\end{thebibliography}
\end{document}